\documentclass[runningheads]{llncs}

\usepackage[colorlinks,urlcolor=blue,linkcolor=blue,citecolor=blue]{hyperref}
\usepackage[T1]{fontenc}
\usepackage{graphicx}
\usepackage{booktabs}       
\usepackage{subcaption}

\graphicspath{{./figures/}}

\def\citep{\cite}
\def\citet{\cite}

\begin{document}
\title{Deep learning in a bilateral brain with hemispheric specialisation}
%
%

\author{Chandramouli Rajagopalan\inst{1}\orcidID{0000-0003-3285-4513} \and 
David Rawlinson\inst{1}\orcidID{0000-0001-9443-3840} \and 
Elkhonon Goldberg\inst{2}\orcidID{0000-0002-4249-1907} \and 
Gideon Kowadlo\inst{1}\orcidID{0000-0001-6036-1180}}
\authorrunning{C. Rajagopalan et al.}
%
\institute{Cerenaut, Australia
\email{\{david,gideon\}@cerenaut.ai}\\
\url{https://cerenaut.ai} \and
Luria Neuroscience Institute, New York, USA
\email{eg@elkhonongoldberg.com}}

\maketitle              
\begin{abstract}
The brains of all bilaterally symmetric animals on Earth are divided into left and right hemispheres. The anatomy and functionality of the hemispheres have a large degree of overlap, but there are asymmetries, and they specialise in possesses different attributes. Other authors have used computational models to mimic hemispheric asymmetries with a focus on reproducing human data on semantic and visual processing tasks. We took a different approach and aimed to understand how dual hemispheres in a bilateral architecture interact to perform well in a given task. We propose a bilateral artificial neural network that imitates lateralisation observed in nature: that the left hemisphere specialises in local features and the right in global features. We used different training objectives to achieve the desired specialisation and tested it on an image classification task with two different CNN backbones: ResNet and VGG. Our analysis found that the hemispheres represent complementary features that are exploited by a network head that implements a type of weighted attention. The bilateral architecture outperformed a range of baselines of similar representational capacity that do not exploit differential specialisation, with the exception of a conventional ensemble of unilateral networks trained on dual training objectives for local and global features. The results demonstrate the efficacy of bilateralism, contribute to the discussion of bilateralism in biological brains, and the principle may serve as an inductive bias for new AI systems.


\end{abstract}

\section{Introduction}

Division of the brain into left and right hemispheres is a remarkably conserved feature of brains across species, suggesting the importance of the bilateral architecture for intelligence.
The anatomy and physiology of the hemispheres have a large degree of overlap, but they are asymmetric and specialise to possess different attributes. 
A likely cause of specialisation is the hemispheric asymmetry, which includes differences in neuron density \cite{xweems_hemispheric_2004,goldberg_hemispheric_2013}, distribution of neurotransmitters \cite{Reggia1998}, firing thresholds \cite{Reggia1998} and intrahemispheric connectivity patterns \cite{Schapiro2013,Peleg2010}.

One of the emergent functional differences is a specialisation for local (left hemisphere) vs global (right hemisphere) features for visual and auditory processing \cite{hsiao_hemispheric_2013}.
In a classic experimental setup \cite{navon_forest_1977}, a large letter is made up of smaller letters. Subjects are faster at detecting the small letters when presented to the left hemisphere, and the large letters when presented to the right.


Computational models that mimic neuroanatomical and physiological asymmetries are typically subjected to standard tasks and compared to human behavioural data. They provide evidence that asymmetries can cause observed lateralisation of activity for specific tasks, e.g., \cite{Lambon2001,Monaghan2008,Schapiro2013,Wang2013}. 
The central question is usually to understand the cause of lateralisation/specialisation.
However, none have been applied to object recognition, and none have tried to understand how differences can be exploited to improve performance on a given task.


The aim of \textit{this} project is to investigate the functional benefit of complementary hemispheres for a given task.
Our objective is twofold. First, improve performance in a vision task using a bilateral architecture.
Second, better understand the functional benefit of a bi-hemispheric architecture and understand how bilateral principles could be exploited more generally for AI/ML.


We focused on the left hemisphere's specialisation in local features and the right's in global features. To investigate this, we chose a hierarchical image classification task where each sample belongs to both a fine and a coarse class, e.g. \textbf{coarse}: sea creature, \textbf{fine}: penguin, seal, shark. 
In this setup, fine labels represent local features while coarse labels represent global features.


Our approach is to construct a bilateral network and induce specialisation in each hemisphere directly, rather than to test if it emerges as a result of architectural asymmetry.
We modelled the hemispheres with left and right CNNs and induced specialisation using supervised training with different objectives, using either coarse or fine labels.
The bilateral network was compared with several baselines that do not have specialisation but do have the same number of learnable parameters. The differences were analysed to explore the effects of bilateral specialisation.
Experiments were repeated with two types of CNN hemispheres, VGG and ResNet variants, to test generalisation over different backbones.

The source code is available at \citep{Rajagopalan2022}. 

\section{Related work}
\label{sec:related_work}

Specialisation for local and global visual features has been attributed to differential sensitivity to spatial frequencies and explored with computational models for face perception \cite{Dailey1999,Hsiao2008,Wang2013} and a line bisection task \cite{Shillcock2001,Monaghan2004b}.
However, other studies do not support hemispheric specialisation for particular frequency ranges, and Hsiao et al. \cite{hsiao_hemispheric_2013} showed that this specialisation may emerge as a result of differences in neuron connectivity.

Beaulieu et al. (2020) \citet{Beaulieu2020} created a dual-stream architecture called Neuromodulator Meta-learner, where one network learns to modulate the other, to enhance continual learning. 
Bakhtiari et al. (2021) \citet{bakhtiari_functional_2021} created a network with parallel pathways to reproduce the functionality of dorsal (`where') and ventral (`what') pathways in an ANN trained with a single loss function.
Li et al. (2021) \citet{Li2021} discussed specialisation of a branched neural network.

In the most related study, Mayan et al. (2021) \citet{Mayan2021} tackled the same classification task with fine and coarse (hierarchical) labels. A single network with two hemispheres was trained with supervised learning and a single loss function. They used different hyperparameters in each hemisphere, with analogies to biological parameters, to encourage specialisation; which was achieved but the bilateral network did not have an advantage over baselines.

Our bilateral approach can be viewed as a type of ensemble, which is a popular approach in ML \cite{Sagi2018}. 
Typically, the outputs of multiple models, trained on different seeds, are combined; and the models are of the same type.
Other techniques encourage diversity within the ensemble using different training objectives, sampling, architecture, or losses \citep{Tian2012,Lee2015,Alam2020}. 
Our work can be viewed as a special case of a diversified ensemble, where the ensemble architecture exploits hierarchical data labels to achieve diversification.

In multi-domain learning (MDL), multiple models are trained to specialise in a particular domain \cite{he_multi-domain_2022}.
Two of the main approaches include a) modelling domain-specific and domain-general features \cite{joshi2012multi} and
b) explicitly modelling the relationship between domains.
MDL is focused on classes from domains with different distributions. 
The domain is sometimes explicit and other times it is determined by the model \cite{chen2018branch}.
In contrast, this project is focused on explicit hierarchical labels from the same distribution. 
The objective being to mimic learning different aspects of a single task with left and right hemispheres. 


\section{Methods}
\label{sec:methods}

\subsection{Dataset}
We used the CIFAR-100 dataset \citep{Krizhevsky2009}, as it includes hierarchical labels that denote fine (local) and coarse (global) classes. There are 100 fine and 20 coarse classes.
The dataset is split into 50,000 training and 10,000 test samples.
Images were resized to 32x32 and we used two widely adopted data augmentation techniques, a random crop and a random horizontal flip (using inbuilt PyTorch functions) to increase diversity of the training set.

\subsection{Models and specialisation approach}
\label{sec:models}
\begin{figure}[t!]
\centering
    \includegraphics[width=1.0\textwidth]{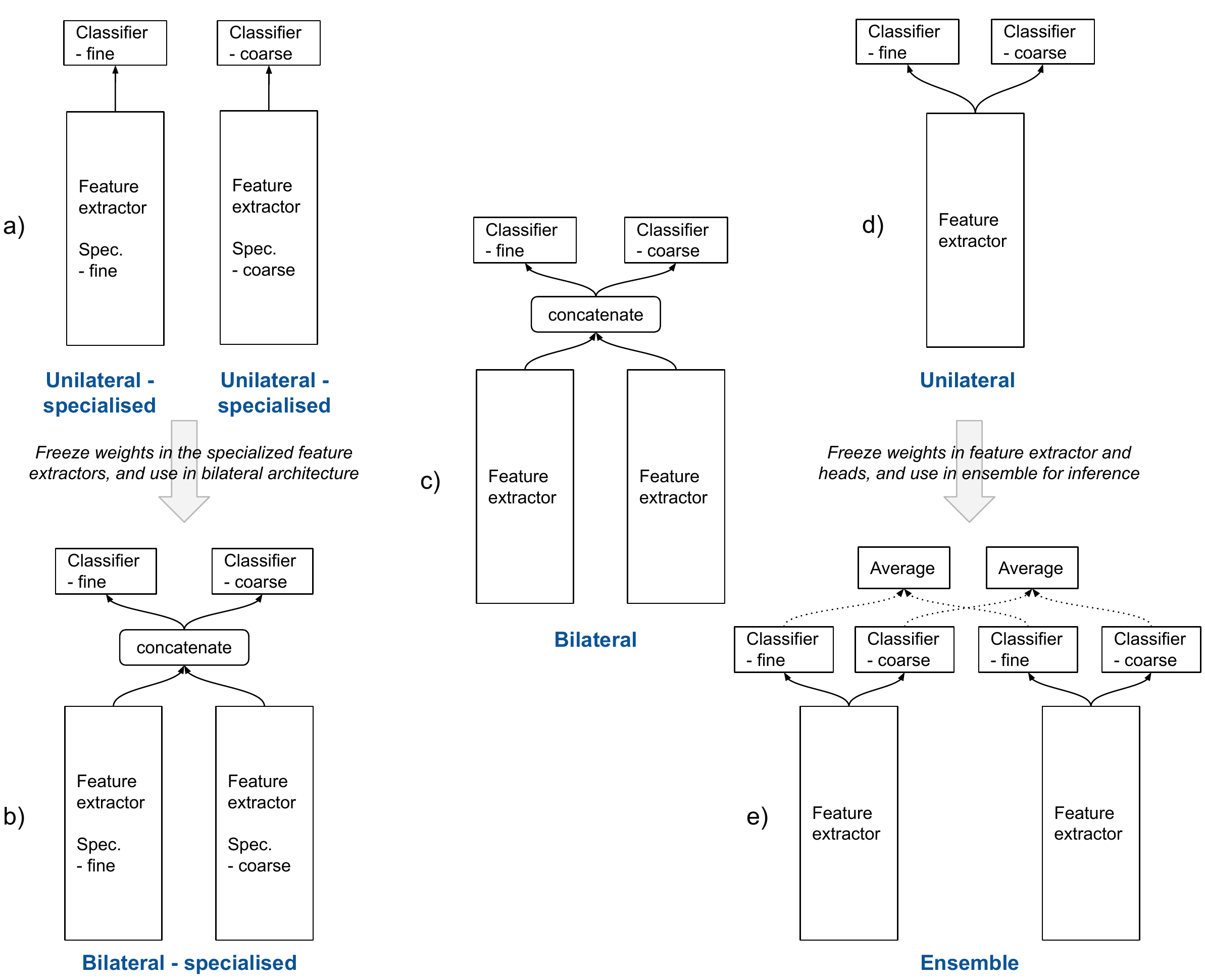}
    \caption{\textbf{Bilateral architecture and baselines.} Feature extractors are either ResNet-9 or VGG-11. The classifier heads are single layer fully connected networks, with output dimensions the same size as the number of classes. For brevity, if not specified, the network is unspecialised.}
    \label{fig:models}
\end{figure}

We compared a bilateral model to several baselines, shown in Figure~\ref{fig:models} and described below.
Details on training and evaluation in Section~\ref{sec:experiments_training}.
Each model consisted of one or more feature extractor, representing hemispheres, as well as one or two classifier heads, one for each label type (fine or coarse).
The hemispheres were implemented with two different CNN architectures to test if the effects generalise across architectures: ResNet \cite{He2015} and VGG \cite{simonyan_very_2015}.
We chose VGG and ResNet because they are common, well understood and amongst the best performing on many vision tasks. 

\subsubsection{Common elements -- backbones and heads}
In the case of ResNet-based models, we empirically optimised the number of layers on the dataset, resulting in a 9-layer network, hereafter referred to as ResNet-9. In shorthand, the full architecture is: \textbf{conv64-conv128-pool-res128,128-pool-conv256-conv512-pool-res512,512-av\_pool}. ConvX = convolutional with X filters, res = residual block with two convolutional blocks of X,Y filters, pool = max pool, av\_pool = average pool. All filters are 3x3 and we used ReLU activation function.

For the VGG model, we chose the VGG-11A variant \cite{simonyan_very_2015}, hereafter referred to as VGG-11, as it is closest in depth to ResNet-9 and works well for CIFAR-100 \cite{nokland_training_2019}.
We used the PyTorch implementation, but removed the final fully-connected layer as we added our own classifier (explained below). In shorthand, the full architecture is: \textbf{conv64-pool-conv128-pool-conv256-conv256-pool-conv512-conv512-pool-conv512-conv512-pool-fc4096-fc4096}. Definitions as above, and fcX = fully connected with X units. All filters are 3x3.
Although VGG-11 and ResNet-9 are both convolutional networks with a similar depth, there are substantial architectural differences. VGG does not have skip connections, and there are a lot more parameters due to the fully-connected layers.

The heads consist of a single fully-connected layer, without bias or non-linearity and with dimensions the same size as the number of classes (10 for coarse and 100 for fine). 

\subsubsection{Bilateral network \textit{with} specialisation}
Assembling and training the network consisted of two phases. 
In Phase 1, individual hemispheres were trained to be specialised feature extractors using supervised learning with a single classifier head, resulting in `Unilateral - specialised', Figure~\ref{fig:models}a. 
The left was specialised for local features using fine labels, and the right was specialised for global features with coarse labels.
In Phase 2, the hemispheres (serving as feature extractors) were brought together into a broader architecture, with two initialised classifier heads, one for fine and coarse labels, resulting in `Bilateral - specialised', Figure~\ref{fig:models}b, .
The weights of the individual hemispheres were frozen and the heads were trained to use the features for classifying fine and coarse labels. 
A single loss was calculated by adding the loss from each head.


\subsubsection{Unilateral network \textit{with} specialisation}
We compared the bilateral network with the individual specialised hemispheres alone, on the tasks for which they were specialised, referred to as Unilateral-specialised, Figure~\ref{fig:models}a. 

\subsubsection{Unilateral network \textit{without} specialisation}
To investigate specialisation, we compared Unilateral-specialised to a single hemisphere \textit{without} specialisation, referred to as Unilateral-unspecialised, Figure~\ref{fig:models}d. 
The single hemisphere was trained on both types of labels simultaneously, using two classifier heads and adding the loss.
This baseline also helps to understand the effect of complementary loss functions; on a single hemisphere compared to the bilateral network.

\subsubsection{Bilateral network \textit{without} specialisation}
To better understand the role of specialisation in the bilateral network, we compared Bilateral-specialised to an equivalent network, with the same number of trainable parameters, but without specialisation, referred to as Bilateral-unspecialised, Figure~\ref{fig:models}c.
It can be viewed as a type of ensemble, where the models are combined in the same way as the specialised bilateral network. 

We trained the entire network (two hemispheres and two heads) without first explicitly inducing specialisation in the individual hemispheres. Like for the unilateral network and bilateral network with specialisation, the loss from each head are added together to create a single loss.

\subsubsection{Ensembles}
To understand the differences between differential specialisation and conventional ensembling, we compared to a 2-model ensemble, Figure~\ref{fig:models}e.
We chose 2 models, so that the ensemble has approximately the same number of trainable parameters as the bilateral network.
We used two unilateral models without specialisation, trained from different seeds. 
The output of the ensemble was the average of the heads: for fine classification, the average of the fine heads, and for coarse, the average of the coarse heads.
We trained five models with different seeds, and then constructed five 2-model ensembles, each with a unique pair of models selected from the pool of five.

\subsection{Training procedure} 
\label{sec:experiments_training}

The models were trained with supervised learning using a cross-entropy loss function. 
In the case of dual heads, the loss from each head was added to create a single loss value.
Weights were initialised with PyTorch defaults, from a random uniform distribution $U(-\sqrt{k},\sqrt{k})$: for linear layers $k=1/in\_features$, and for convolutional layers $k = \frac{1}{C_{in} * \Pi_{i=0}^{1} kernel\_size[i]}$.
We used weight decay of 1.0e-5, dropout of 0.6 on the classifier heads 
to regularise the models, Adam optimiser \citep{Kingma2014}, a learning rate of 1.0e-4 and mini-batch size of 256. These hyperparameters were empirically optimised with manual coordinate descent.

Each network was trained for 180 epochs from 5 random seeds. At the end of each epoch, the models were validated on the test split. 
The number of epochs was sufficient to ensure that the validation accuracy had plateaued, and the final model was the one with the highest validation accuracy across epochs.

\subsection{Visualising network operation}
\label{sec:experiments_vis}

We focused on one backbone, ResNet-9, and used two types of visualisations to better understand the operation of the bilateral network.
To be informative, we selected key scenarios that are distinct from each other and highlight the contribution of different parts of the network, described below. The definition of `correct' for the bilateral-specialised network, is that it was successful for both fine and coarse class labels. 
\begin{itemize}
	\item Scenario 1: Bilateral-specialised network is correct, left and right hemispheres are incorrect
	\item Scenario 2: Bilateral-specialised network and right hemisphere are correct, left hemisphere is incorrect
	\item Scenario 3: Bilateral-specialised network and left hemisphere are correct, right hemisphere is incorrect
	\item Scenario 4: Bilateral-specialised network is incorrect, left and right hemispheres are correct
\end{itemize}

\subsubsection{Gradient camera (Grad-Cam) visualisation}
To understand how the specialised bilateral network exploits features from both hemispheres, we visualised gradient flow with respect to target convolutional layers \citep{Selvaraju2016,Chattopadhyay2017} using the Grad-Cam library \cite{Gildenblat2021}. The gradients reveal which areas of the image contribute most to the prediction; the gradient heatmap highlights the region of focus. We applied the visualisation to each of the specialised hemispheres individually, and compared it to the bilateral network.

For a single hemisphere, the gradients were averaged over the 2nd convolutional layer of each residual block (2 convolutional layers with skip connections) during prediction. In the case of the bilateral network, the gradients were averaged over both hemispheres during simultaneous prediction of both coarse and fine labels (the network had two heads).

\subsubsection{Feature analysis using cosine similarity}
We investigated how left and right features are exploited by the bilateral network by analysing the relationship between representations in different parts of the network, with a focus on how the representations are transformed by the network heads.
We did this by measuring the similarity of features for images of the same label. They are expected to have similar features, so measuring the feature similarity at different parts of the network should be revealing.

We first grouped the image samples into random pairs with the same class label. We then plotted a bivariate distribution of cosine similarity between the pairs, one for the input to the heads, denoted `concatenated' and the other for the average of the network head outputs, denoted `bilateral'.
Similarity is plotted along the following dimensions: left hemisphere, right hemisphere and a third dimension, either concatenated or bilateral.
In addition, univariate marginal distributions were plotted for each hemisphere.

\subsection{Experimental setup} 
Experiments were conducted with the PyTorch-Lightning \citep{Falcon2019} research framework, a wrapper around the PyTorch library \citep{Paszke2019}.
All models were trained and evaluated with a virtual machine with a single NVIDIA A10 GPU, 24GB RAM (through Lambda Labs \url{https://lambdalabs.com/}).

\section{Results}
\label{sec:results}

\subsection{Accuracy}

\begin{figure*}[thb!]
    \includegraphics[width=1\textwidth]{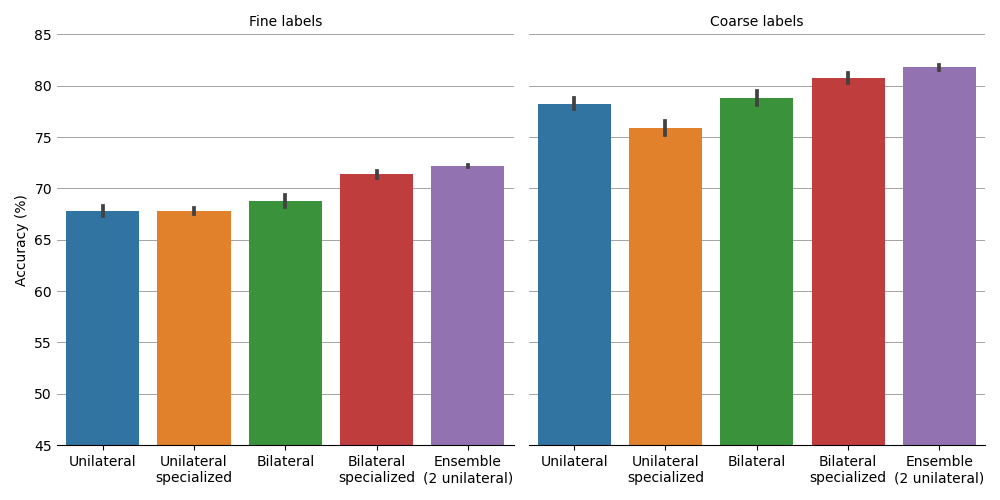}
    \caption{\textbf{Accuracy of ResNet-based models.} For brevity, if not specified, the network is unspecialised.}
    \label{fig:results_resnet}
\end{figure*}
\begin{figure*}
    \includegraphics[width=1\textwidth]{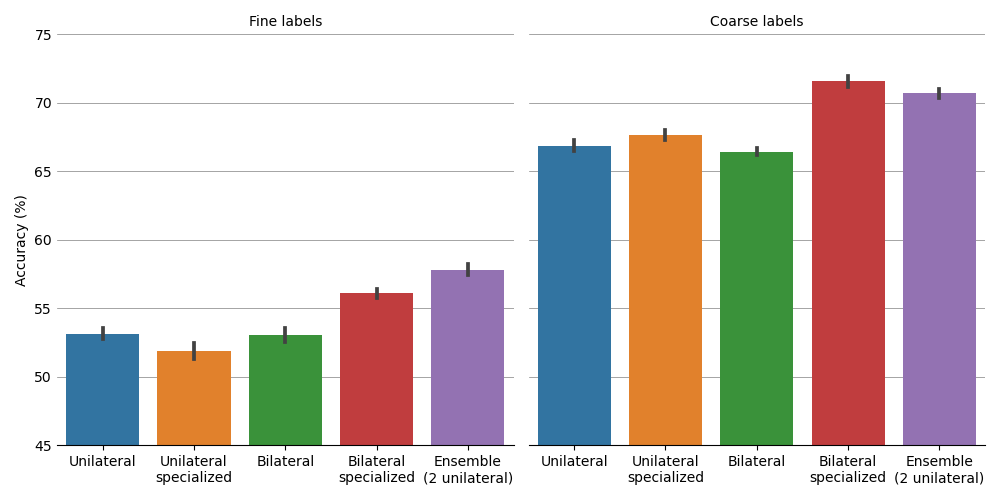}
    \caption{\textbf{Accuracy of VGG-11-based models.} For brevity, if not specified, the network is unspecialised.}
    \label{fig:results_vgg}
\end{figure*}

The relative performance of models was very consistent between ResNet and VGG backbones, Tables~\ref{table:results_resnet} and \ref{table:results_vgg} and Figures~\ref{fig:results_resnet} and \ref{fig:results_vgg}. Accuracy given as mean $\pm$ 1 standard deviation (shown as error bars in the plots) of 5 seeds.
In summary, Bilateral-specialised outperformed all the baselines, except for the 2-model ensemble.
The 2-model ensemble consists of unilateral models, which were more effective than specialised unilateral models in most cases.
The exception was the VGG backbone with coarse labels, where the unspecialised unilateral model \textit{was not} better than the specialised version, and the resulting 2-model ensemble did not outperform Bilateral-specialised.



\begin{table*}[ht!]
  \centering
  \begin{tabular}{cccc}
    \toprule
    {\textbf {Model}} & {\textbf {\# Params (M)}} & \begin{tabular}{c}\textbf{Accuracy (\%)} \\ \textbf{- fine labels} \end{tabular} & \begin{tabular}{c}\textbf{Accuracy (\%)} \\ \textbf{- coarse labels} \end{tabular}\\    
    \midrule
	Unilateral & 6.63 & 67.81 $\pm$ 0.46 & 78.25 $\pm$ 0.51 \\
    Unilateral-specialised & 6.58 & 67.81 $\pm$ 0.31 & \textemdash \\
    Unilateral-specialised & 6.62 & \textemdash & 75.87 $\pm$ 0.70 \\
    \midrule
	Bilateral & 13.26 & 68.80 $\pm$ 0.59 & 78.82 $\pm$ 0.65 \\
	Bilateral-specialised & 13.26 & 71.35 $\pm$ 0.36 & 80.71 $\pm$ 0.48 \\
    \midrule
	Ensemble (2 unilateral models) & 13.26 & 72.15 $\pm$ 0.11 & 81.80 $\pm$ 0.24 \\
    \bottomrule
  \end{tabular}
    \caption{Accuracy of ResNet-based models.}
    \label{table:results_resnet}
\end{table*}

\begin{table*}[ht!]
  \centering
  \begin{tabular}{cccc}
    \toprule
    {\textbf {Model}} & {\textbf {\# Params (M)}} & \begin{tabular}{c}\textbf{Accuracy (\%)} \\ \textbf{- fine labels} \end{tabular} & \begin{tabular}{c}\textbf{Accuracy (\%)} \\ \textbf{- coarse labels} \end{tabular}\\
    \midrule
    Unilateral & 129.26 & 53.13 $\pm$ 0.41 & 66.86 $\pm$ 0.41 \\
	Unilateral-specialised & 129.18 & 51.87 $\pm$ 0.58 & \textemdash \\
	Unilateral-specialised & 128.85 & \textemdash & 66.86 $\pm$ 0.41 \\
    \midrule
	Bilateral & 258.52 & 53.05 $\pm$ 0.49 & 66.42 $\pm$ 0.27 \\
	Bilateral-specialised & 258.52  & 56.09 $\pm$ 0.33 & 71.57 $\pm$ 0.42 \\
    \midrule
	Ensemble (2 unilateral models) & 258.52 & 57.81 $\pm$ 0.38 & 70.68 $\pm$ 0.30 \\
    \bottomrule
  \end{tabular}
    \caption{Accuracy of VGG-based models}
    \label{table:results_vgg}
\end{table*}

\subsection{Visualising network operation}

\begin{figure*}[thb!]
    \begin{subfigure}{0.24\textwidth}
    \centering
        \includegraphics[width=1.5cm, height=1.5cm]{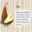}
        \caption{Bosc Image}
        \label{fig:grad_cam_original_bosc}
    \end{subfigure}
    \hfill
    \begin{subfigure}{0.24\textwidth}
    \centering
        \includegraphics[width=1.5cm, height=1.5cm]{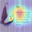}
        \caption{Left}
        \label{fig:grad_cam_narrow_bosc}
    \end{subfigure}    
    \hfill
    \begin{subfigure}{0.24\textwidth}
    \centering
        \includegraphics[width=1.5cm, height=1.5cm]{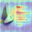}
        \caption{Right}
        \label{fig:grad_cam_broad_bosc}
    \end{subfigure}
    \hfill
    \begin{subfigure}{0.24\textwidth}
    \centering
        \includegraphics[width=1.5cm, height=1.5cm]{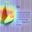}
        \caption{Bilateral-sp}
        \label{fig:grad_cam_bilateral_bosc}
    \end{subfigure}
    
    \caption{\textbf{Grad-Cam for Scenario 1: The bilateral-specialised (abbreviated to bilateral-sp in the figure) network is correct, left and right are incorrect.} The bilateral network appears to adjust the area of focus even when both the hemispheres' features are situated external to the bosc.}
    \label{fig:grad_cam_visualizations_a}
\end{figure*}
\begin{figure*}[thb!]
    \begin{subfigure}{0.24\textwidth}
    \centering
        \includegraphics[width=1.5cm, height=1.5cm]{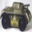}
        \caption{Tank Image}
        \label{fig:grad_cam_original_tank}
    \end{subfigure}
    \hfill
    \begin{subfigure}{0.24\textwidth}
    \centering
        \includegraphics[width=1.5cm, height=1.5cm]{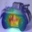}
        \caption{Left}
        \label{fig:grad_cam_narrow_tank}
    \end{subfigure}
    \hfill
    \begin{subfigure}{0.24\textwidth}
    \centering
        \includegraphics[width=1.5cm, height=1.5cm]{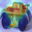}
        \caption{Right}
        \label{fig:grad_cam_broad_tank}
    \end{subfigure}
    \hfill
    \begin{subfigure}{0.24\textwidth}
    \centering
        \includegraphics[width=1.5cm, height=1.5cm]{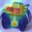}
        \caption{Bilateral-sp}
        \label{fig:grad_cam_bilateral_tank}
    \end{subfigure}
    
    \caption{\textbf{Grad-Cam for Scenario 2: The bilateral-specialised network and the right are correct, the left is incorrect.} Possibly due to the fact that there are multiple labels with wheels, the left failed to differentiate the tank by its wheels alone, however the bilateral model appears to overcome this, using the right's features.}
    \label{fig:grad_cam_visualizations_b}
\end{figure*}
\begin{figure*}[thb!]
    \begin{subfigure}{0.24\textwidth}
    \centering
        \includegraphics[width=1.5cm, height=1.5cm]{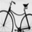}
        \caption{Cycle Image}
        \label{fig:grad_cam_original_cycle}
    \end{subfigure}
    \hfill
    \begin{subfigure}{0.24\textwidth}
    \centering
        \includegraphics[width=1.5cm, height=1.5cm]{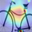}
        \caption{Left}
        \label{fig:grad_cam_narrow_cycle}
    \end{subfigure}    
    \hfill
    \begin{subfigure}{0.24\textwidth}
    \centering
        \includegraphics[width=1.5cm, height=1.5cm]{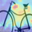}
        \caption{Right}
        \label{fig:grad_cam_broad_cycle}
    \end{subfigure}
    \hfill
    \begin{subfigure}{0.24\textwidth}
    \centering
        \includegraphics[width=1.5cm, height=1.5cm]{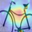}
        \caption{Bilateral-sp}
        \label{fig:grad_cam_bilateral_cycle}
    \end{subfigure}
  
    \caption{\textbf{Grad-Cam for Scenario 3: The bilateral-specialised network and the left are correct, the right is incorrect.} The left hemisphere identified more local features of the bicycle compared to the right. The network head appear to exploit the left representations to make an accurate prediction.}
    \label{fig:grad_cam_visualizations_c}
\end{figure*}
\begin{figure*}[thb!]
    \begin{subfigure}{0.24\textwidth}
    \centering
        \includegraphics[width=1.5cm, height=1.5cm]{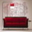}
        \caption{Sofa Image}
        \label{fig:grad_cam_original_sofa}
    \end{subfigure}
    \hfill
    \begin{subfigure}{0.24\textwidth}
    \centering
        \includegraphics[width=1.5cm, height=1.5cm]{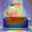}
        \caption{Left}
        \label{fig:grad_cam_narrow_sofa}
    \end{subfigure}
	\hfill
    \begin{subfigure}{0.24\textwidth}
    \centering
        \includegraphics[width=1.5cm, height=1.5cm]{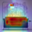}
        \caption{Right}
        \label{fig:grad_cam_broad_sofa}
    \end{subfigure}    
    \hfill
    \begin{subfigure}{0.24\textwidth}
    \centering
        \includegraphics[width=1.5cm, height=1.5cm]{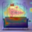}
        \caption{Bilateral-sp}
        \label{fig:grad_cam_bilateral_sofa}
    \end{subfigure}

    \caption{\textbf{Grad-Cam for Scenario 4: The bilateral-specialised network is incorrect, the left and right hemispheres are correct.} Both hemispheres identified local and global features. The network heads appear to over-compensate and focus on the region outside of the sofa, which is ultimately unsuccessful.}
    \label{fig:grad_cam_visualizations_d}
\end{figure*}

Grad-Cam visualisations are shown in Figures~\ref{fig:grad_cam_visualizations_a} to \ref{fig:grad_cam_visualizations_d} for scenarios described in Section~\ref{sec:experiments_vis}.
In general, the features appear to be more local in the left hemisphere and more global in the right. 
The heads blend different aspects of features from the left and right, for the task.
The effect is that in many cases, \textit{features trained for fine classes are helpful for coarse classes and vice versa, and a correct classification can be achieved even if both hemispheres are individually incorrect}.

\begin{figure*}[thb!]
\centering
    \begin{subfigure}{0.45\textwidth}
    \centering
        \includegraphics[width=1.\textwidth]{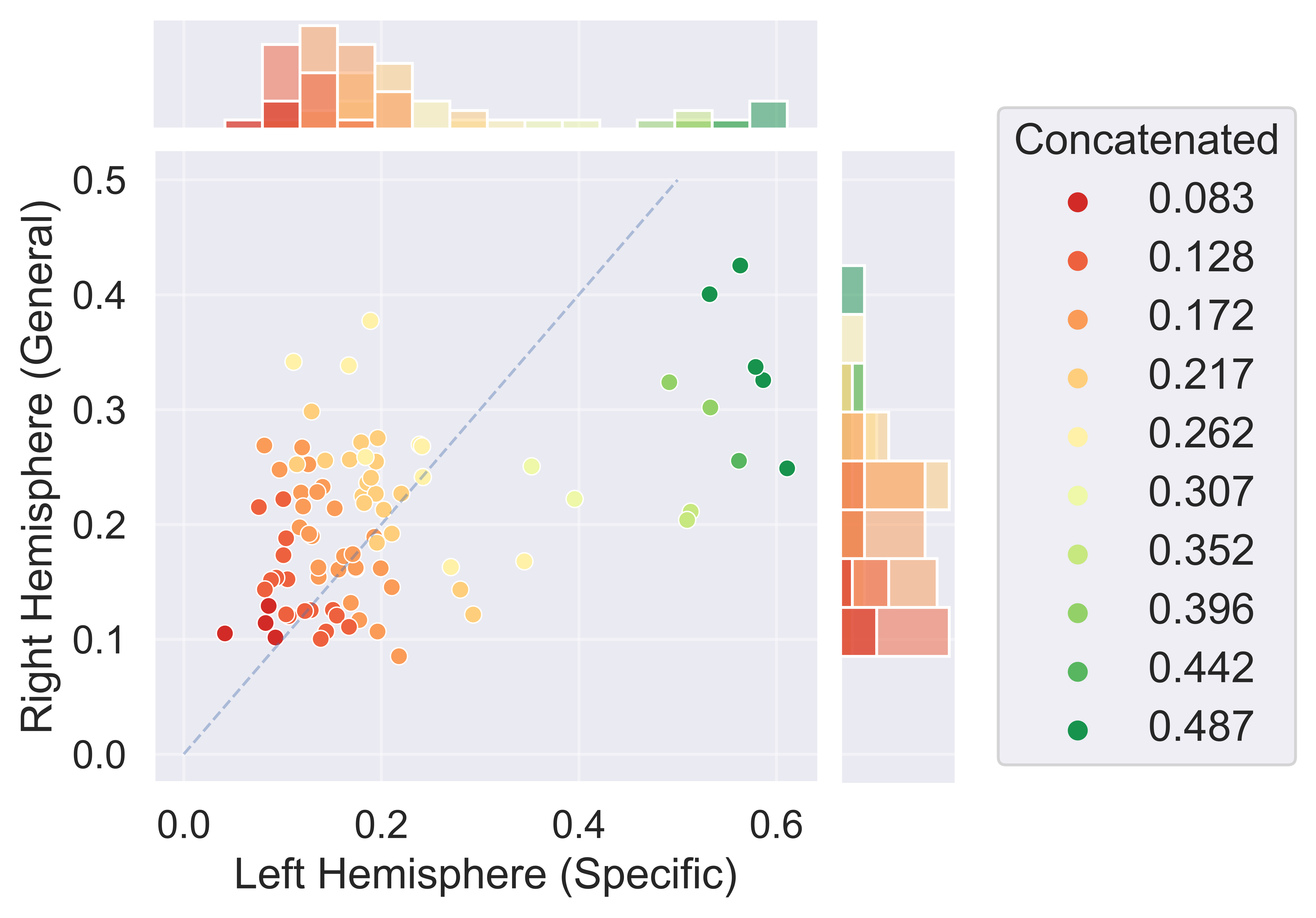}
        \caption{Concatenated}
        \label{fig:cosine_unspecial_nfbfbicamntbt}
    \end{subfigure}
    \hfill
    \begin{subfigure}{0.45\textwidth}
    \centering
        \includegraphics[width=1.\textwidth]{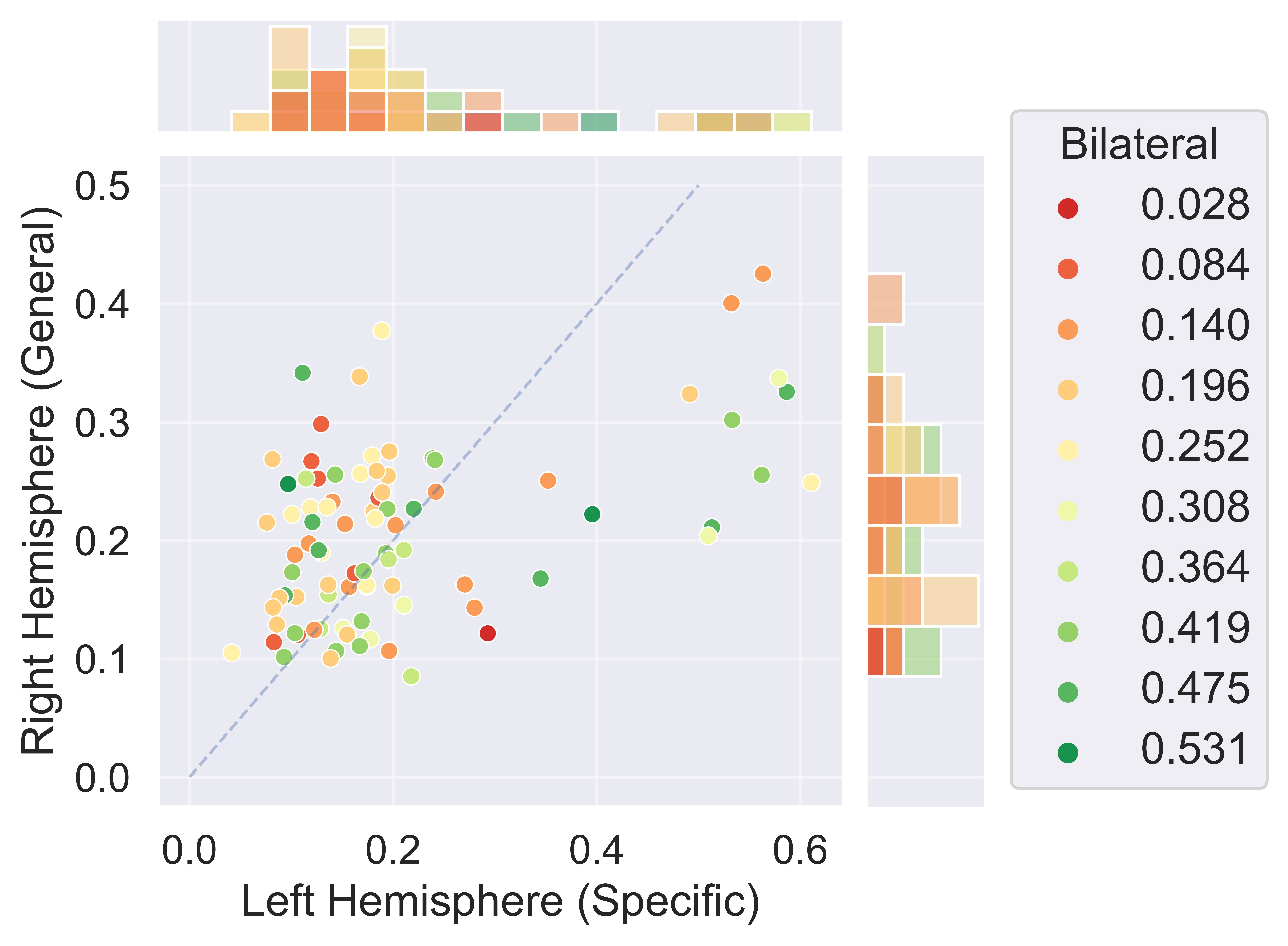}
        \caption{Bilateral-specialised network}
        \label{fig:cosine_special_nfbfbicamntbt}
    \end{subfigure}
    \caption{\textbf{Cosine similarity distribution for Scenario 1: The bilateral-specialised network is correct, left and right are incorrect.} Many of the pairs (same label) have dissimilar (values closer to 0) features in left and right, seen in the density of points close to the origin. The concatenated features are also dissimilar, but the bilateral network elevates the similarity of many of the points. Each point is a pair of images with the same label. The x-axis is similarity in the left hemisphere, y-axis is similarity in the right hemisphere. The color is similarity in the combined representation. Additionally, the univariate marginal distributions are shown for left and right with histograms above and to the side of the bivariate distribution.}
    \label{fig:cosine_analysis_a}
\end{figure*}

\begin{figure*}[thb!]
\centering
    \begin{subfigure}{0.45\textwidth}
    \centering
        \includegraphics[width=1.\textwidth]{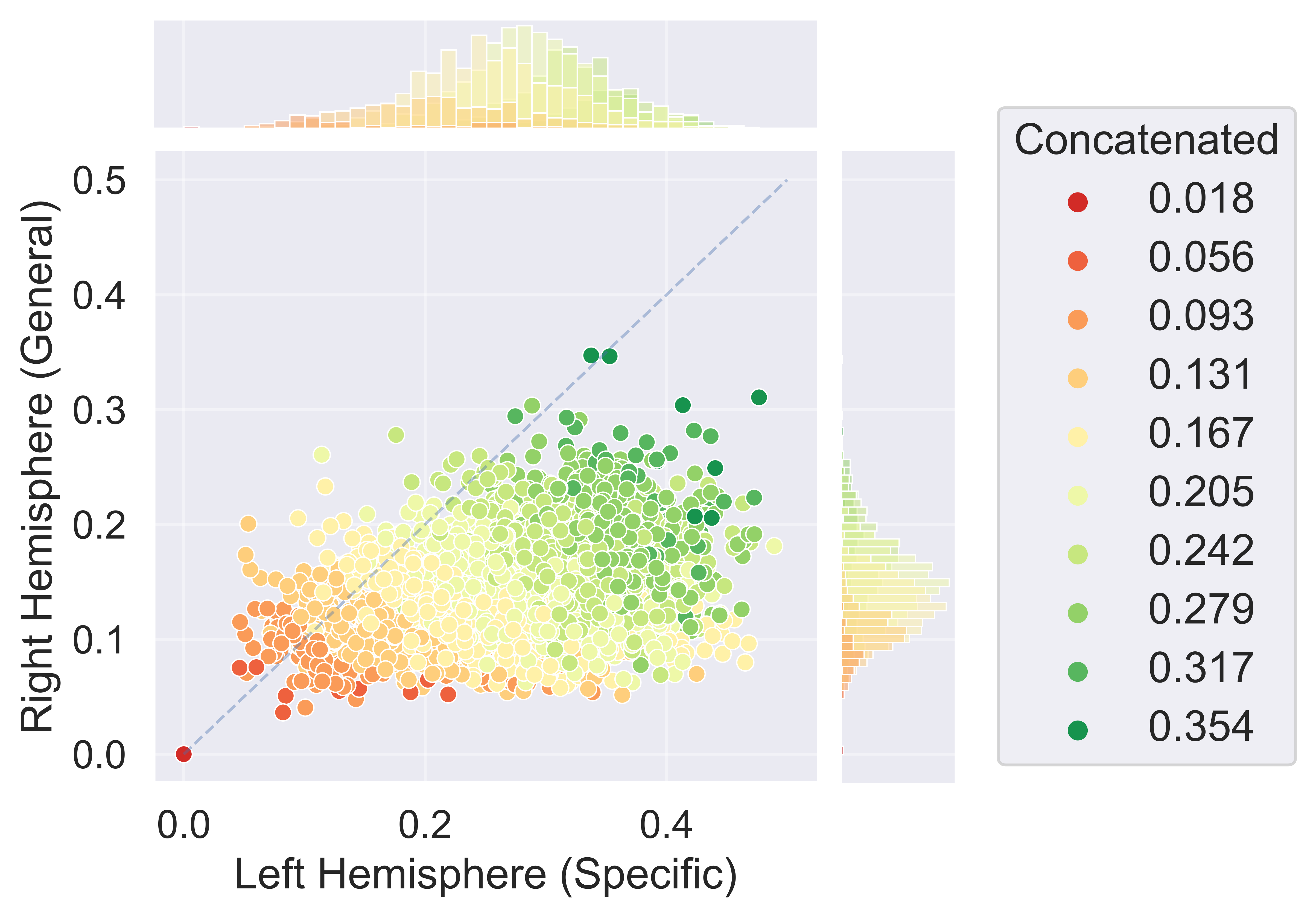}
        \caption{Concatenated}
        \label{fig:cosine_unspecial_nfbtbicamntbt}
    \end{subfigure}
    \hfill
    \begin{subfigure}{0.45\textwidth}
    \centering
        \includegraphics[width=1.\textwidth]{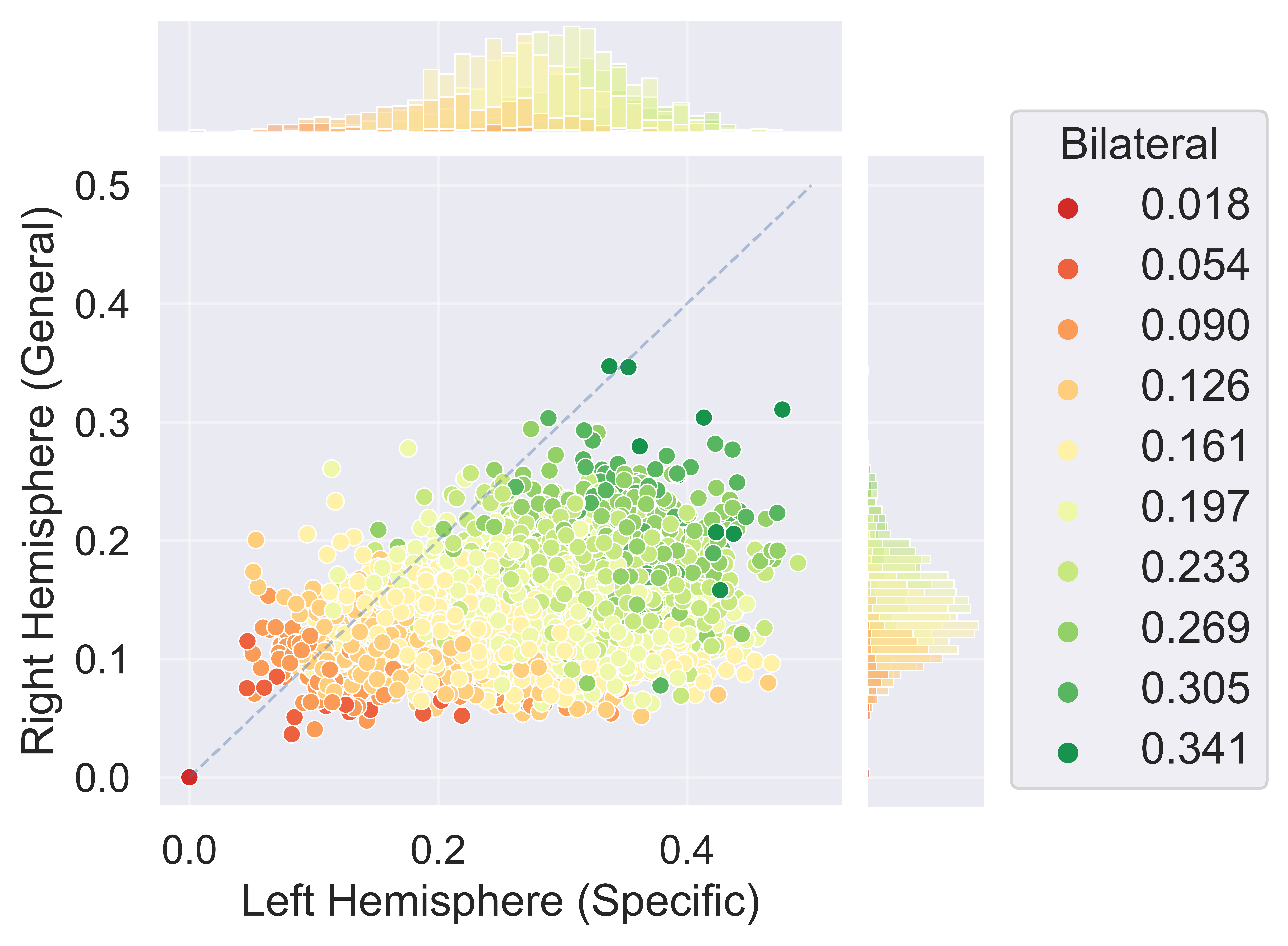}
        \caption{Bilateral-specialised network}
        \label{fig:cosine_special_nfbtbicamntbt}
    \end{subfigure}    
    \caption{\textbf{Cosine similarity distribution for Scenario 2: The bilateral-specialised network and the right are correct, the left is incorrect.} Some regions with points that have low similarity in the left (at approximately x, y = 0.2, 0.2), have elevated similarity in the bilateral output, compared to concatenated. Also, for points with high similarity in the left (approximately 0.35, 0.1), the similarity is lower in the bilateral output. In other words, the bilateral network was able to exploit features in the right hemisphere over the left. Each point is a pair of images with the same label. The x-axis is similarity in the left hemisphere, y-axis is similarity in the right hemisphere. The color is similarity in the combined representation. Additionally, the univariate marginal distributions are shown for left and right with histograms above and to the side of the bivariate distribution.}
    \label{fig:cosine_analysis_b}
\end{figure*}

The cosine similarity results are shown in Figure~\ref{fig:cosine_analysis_a} and Figure\ref{fig:cosine_analysis_b}.
In the concatenated distribution, there is a strong correlation between similarity in left and concatenated, and right and concatenated.
In contrast, there is no obvious correlation in the bilateral distribution.
The network heads appear to have learned a non-linear transformation of the feature space.

\section{Discussion}


We found that there is an advantage of two hemispheres over one, demonstrated by the fact that the specialised bilateral architecture (Bilateral-specialised) outperformed individual hemispheres (Unilateral-specialised and Unilateral).
Furthermore, we found that it is not due to the greater number of learnable parameters, and having two hemispheres is not sufficient, but that specialisation is important, shown by the benefit of Bilateral-specialised over Bilateral.

Surprisingly, unspecialised hemispheres (Unilateral) performed as well as or better than specialised hemispheres (Unilateral-specialised) in most cases.
Recall that unilateral models are trained with two objective functions (fine and coarse). 
So using two training objectives led to better performance on both label types, whether that is in a bilateral or unilateral architecture. 
Using two explicit objectives could be seen as a form of specialisation, even when used simultaneously and in a single model.

Ultimately, two hemispheres arranged in a conventional ensemble achieved the best results.
A key difference with Bilateral-specialised is that the component models of the ensemble were more effective to begin with; benefiting from two training objectives.
Another difference is the way that the model outputs are combined -- averaging of the classification vs concatenation of the feature vector. Future work could explore this in more detail.

Both the specialised bilateral network and the ensemble were able to boost performance of individual hemispheres. 
However, they use different mechanisms.
The novel finding of the paper is that a bilateral architecture can exploit specialised hemispheres, and the remainder of the paper examines this principle.

\subsubsection{Computational cost}
The performance boost of bilateral models over a single hemisphere may appear modest relative to the additional computational cost, which could limit their practical applications. However, incremental gains are difficult to achieve at high levels of performance, making the ~2.5-5\% improvement (across fine, coarse and hemisphere architectures) consequential. Like other new architectures, improvements can be expected in the future.

\subsubsection{How does the bilateral architecture work?}
The Grad-Cam images reveal that the left hemisphere extracts more localised features than the right. 
Different learning objectives enable them to capture different aspects of the environment.
Collectively, the set of features is greater than one network with one objective.
Interestingly, even though the left is explicitly trained on fine labels, the features it extracts are helpful for coarse classes. The inverse is also true.

With reference to the cosine similarity visualisation, before the network heads, the similarity of pairs of images of the same class is correlated with left and right and the similarity is no longer correlated after the network heads; which suggests that the heads implement a non-linear transformation.
In many cases, when left or right produce ineffective features, shown by low similarity between embeddings of these images of the same class, the heads are able to produce features that have increased similarity.
The network heads learn to combine the features from left and right, to different degrees, to produce better predictions. In some cases, the bilateral network makes a correct classification, even if both hemispheres are individually wrong.

In summary, specialisation creates a higher diversity of features.
The network heads implement a type of weighted attention to left and right hemispheres selectively in a task-dependent manner, improving overall class prediction.

\subsubsection{Relation to biology}
A limitation in terms of biological accuracy was that interhemispheric connectivity was achieved at the outputs of the hemispheres. In contrast, biological hemispheres are interconnected throughout their hierarchies \cite{Carson2020}. However, heads may serve a similar purpose, albeit in a cruder way. Interhemispheric connectivity comprises a complex combination of inhibitory and excitatory projections. The hemisphere that is better able to represent the input is likely to have stronger activation and thus inhibit the other hemisphere \citep{weems_hemispheric_2004}. Like the heads, this too is a type of selectivity. Interconnected hemispheres throughout the hierarchy is a topic for future work.

The main finding, that left and right hemispheres extracted local and non-local features to improve performance on a given task, provides one plausible reason (and testable hypothesis) for the prevalence of bilateral brains in nature.
A possible approach to investigate this is with transcranial magnetic stimulation (TMS) to selectively impair one hemisphere at a time on controlled tasks \cite{Pobric2008}.

\section{Conclusion}

Inspired by our bilateral brains, we built a bilateral architecture with left and right neural networks using two CNN backbones and compared them to baselines that captured distinct characteristics of the proposed bilateral network.
We used a classification task with hierarchical classes that captured an observed characteristic of biological hemispheres, that the left is more specialised for fine classes and the right for coarse classes.
The hemispheres were trained to specialise on fine and coarse classes accordingly.
They extracted specialised features and through a type of `weighted attention' by a simple fully connected layer, outperformed various (but not all) baselines on classification of both fine and coarse classes. 
The specialised representations had benefits above the explicit objective of their individual hemispheres.
The results demonstrate that small procedural changes to training can achieve specialisation, and that specialisation can be complementary and beneficial for certain tasks.
The operation of the artificial network provides ideas for the study of neurobiological hemispheric specialisation.
Simultaneously, this work shows that neuroscientific principles can provide inductive biases for novel ML architectures. Currently, in AI/ML where scaling of existing architectures is achieving great success, it's interesting to look at new principles on smaller architecture that could then be scaled.

\begin{credits}
\subsubsection{\ackname} Thanks to Punarjay Chakravarty for helpful discussions.
\end{credits}


\bibliographystyle{splncs04}
\bibliography{references}

\begin{thebibliography}{10}
\providecommand{\url}[1]{\texttt{#1}}
\providecommand{\urlprefix}{URL }
\providecommand{\doi}[1]{https://doi.org/#1}

\bibitem{Alam2020}
Alam, K.M.R., Siddique, N., Adeli, H.: A dynamic ensemble learning algorithm for neural networks. Neural Computing and Applications  \textbf{32},  8675--8690 (6 2020)

\bibitem{bakhtiari_functional_2021}
Bakhtiari, S., Mineault, P., Lillicrap, T., Pack, C., Richards, B.: The functional specialization of visual cortex emerges from training parallel pathways with self-supervised predictive learning. In: Advances in {Neural} {Information} {Processing} {Systems}. vol.~34, pp. 25164--25178. Curran Associates, Inc. (2021)

\bibitem{Beaulieu2020}
Beaulieu, S., Frati, L., Miconi, T., Lehman, J., Stanley, K.O., Clune, J., Cheney, N.: Learning to {Continually} {Learn}. In: \{{ECAI}\} 2020 - 24th {European} {Conference} on {Artificial} {Intelligence} (2020)

\bibitem{Carson2020}
Carson, R.G.: Inter-hemispheric inhibition sculpts the output of neural circuits by co-opting the two cerebral hemispheres. The Journal of Physiology  \textbf{598},  4781--4802 (11 2020)

\bibitem{Chattopadhyay2017}
Chattopadhay, A., Sarkar, A., Howlader, P., Balasubramanian, V.N.: Grad-cam++: Generalized gradient-based visual explanations for deep convolutional networks. In: 2018 IEEE Winter Conference on Applications of Computer Vision (WACV). pp. 839--847 (2018)

\bibitem{chen2018branch}
Chen, Y., Lu, R., Zou, Y., Zhang, Y.: Branch-activated multi-domain convolutional neural network for visual tracking. Journal of Shanghai Jiaotong University (Science)  \textbf{23},  360--367 (2018)

\bibitem{Dailey1999}
Dailey, M.N., Cottrell, G.W.: Organization of face and object recognition in modular neural network models. Neural Networks  \textbf{12},  1053--1074 (10 1999)

\bibitem{Falcon2019}
Falcon, W., contributors: Pytorch lightning (v1.4.1) [computer software] (2019), \url{https://github.com/PyTorchLightning/pytorch-lightning}

\bibitem{Gildenblat2021}
Gildenblat, J., contributors: Pytorch library for cam methods (v1.4.5) [computer software] (2021), \url{https://github.com/jacobgil/pytorch-grad-cam}

\bibitem{goldberg_hemispheric_2013}
Goldberg, E., Roediger, D., Kucukboyaci, N.E., Carlson, C., Devinsky, O., Kuzniecky, R., Halgren, E., Thesen, T.: Hemispheric asymmetries of cortical volume in the human brain. Cortex  \textbf{49}(1),  200--210 (Jan 2013)

\bibitem{He2015}
He, K., Zhang, X., Ren, S., Sun, J.: Deep residual learning for image recognition. Proceedings of the IEEE Computer Society Conference on Computer Vision and Pattern Recognition pp. 770--778 (9 2015)

\bibitem{he_multi-domain_2022}
He, R., Liu, S., He, S., Tang, K.: Multi-domain active learning: Literature review and comparative study. IEEE Transactions on Emerging Topics in Computational Intelligence  \textbf{7}(3),  791--804 (2023)

\bibitem{hsiao_hemispheric_2013}
Hsiao, J.H., Cipollini, B., Cottrell, G.W.: Hemispheric {Asymmetry} in {Perception}: {A} {Differential} {Encoding} {Account}. Journal of Cognitive Neuroscience  \textbf{25}(7),  998--1007 (Jul 2013)

\bibitem{Hsiao2008}
Hsiao, J.H.W., Shieh, D.X., Cottrell, G.W.: Convergence of the visual field split: Hemispheric modeling of face and object recognition. Journal of Cognitive Neuroscience  \textbf{20},  2298--2307 (12 2008)

\bibitem{joshi2012multi}
Joshi, M., Dredze, M., Cohen, W., Rose, C.: Multi-domain learning: when do domains matter? In: Proceedings of the 2012 Joint Conference on Empirical Methods in Natural Language Processing and Computational Natural Language Learning. pp. 1302--1312 (2012)

\bibitem{Kingma2014}
Kingma, D.P., Ba, J.: Adam: A method for stochastic optimization. arXiv preprint arXiv:1412.6980  (2014)

\bibitem{Krizhevsky2009}
Krizhevsky, A., Hinton, G.: Learning multiple layers of features from tiny images. Tech. rep., University of Toronto (2009), publisher: Toronto, ON, Canada

\bibitem{Lee2015}
Lee, S., Purushwalkam, S., Cogswell, M., Crandall, D., Batra, D.: Why m heads are better than one: Training a diverse ensemble of deep networks. arXiv preprint arXiv:1511.06314  (2015)

\bibitem{Li2021}
Li, C., Deza, A.: What matters in branch specialization? using a toy task to make predictions. In: SVRHM 2021 Workshop@ NeurIPS (2021)

\bibitem{Mayan2021}
Mayan, A., Kowadlo, G., Kuhlmann, L.: Right and Left Neural Networks – Inspired by the bicameral brain. Master's thesis, Monash University (2021)

\bibitem{Monaghan2004b}
Monaghan, P., Shillcock, R.: Hemispheric asymmetries in cognitive modeling: connectionist modeling of unilateral visual neglect. Psychological review  \textbf{111},  283--308 (4 2004)

\bibitem{Monaghan2008}
Monaghan, P., Shillcock, R.: Hemispheric dissociation and dyslexia in a computational model of reading. Brain and Language  \textbf{107},  185--193 (12 2008)

\bibitem{navon_forest_1977}
Navon, D.: Forest before trees: {The} precedence of global features in visual perception. Cognitive Psychology  \textbf{9}(3),  353--383 (Jul 1977)

\bibitem{nokland_training_2019}
N{\o}kland, A., Eidnes, L.H.: Training neural networks with local error signals. In: International conference on machine learning. pp. 4839--4850. PMLR (2019)

\bibitem{Paszke2019}
Paszke, A., Gross, S., Massa, F., Lerer, A., Bradbury, J., Chanan, G., Killeen, T., Lin, Z., Gimelshein, N., Antiga, L., Desmaison, A., Köpf, A., Yang, E., DeVito, Z., Raison, M., Tejani, A., Chilamkurthy, S., Steiner, B., Fang, L., Bai, J., Chintala, S.: Pytorch: An imperative style, high-performance deep learning library. Advances in Neural Information Processing Systems  \textbf{32} (2019)

\bibitem{Peleg2010}
Peleg, O., Manevitz, L., Hazan, H., Eviatar, Z.: Two hemispheres—two networks: a computational model explaining hemispheric asymmetries while reading ambiguous words. Annals of Mathematics and Artificial Intelligence 2010 59:1  \textbf{59},  125--147 (8 2010)

\bibitem{Pobric2008}
Pobric, G., Mashal, N., Faust, M., Lavidor, M.: The role of the right cerebral hemisphere in processing novel metaphoric expressions: A transcranial magnetic stimulation study. Journal of Cognitive Neuroscience  \textbf{20},  170--181 (1 2008)

\bibitem{Rajagopalan2022}
Rajagopalan, C., Kowadlo, G.: Bilateral brain (v1.0) [computer software] (2022), \url{https://github.com/Cerenaut/bilateral-brain}

\bibitem{Lambon2001}
Ralph, M.A.L., Mcclelland, J.L., Patterson, K., Galton, C.J., Hodges, J.R.: No right to speak? the relationship between object naming and semantic impairment:neuropsychological evidence and a computational model. Journal of Cognitive Neuroscience  \textbf{13},  341--356 (4 2001)

\bibitem{Reggia1998}
Reggia, J.A., Goodall, S., Shkuro, Y.: Computational studies of lateralization of phoneme sequence generation. Neural Computation  \textbf{10},  1277--1297 (7 1998)

\bibitem{Sagi2018}
Sagi, O., Rokach, L.: Ensemble learning: A survey. Wiley Interdisciplinary Reviews: Data Mining and Knowledge Discovery  \textbf{8},  e1249 (7 2018)

\bibitem{Schapiro2013}
Schapiro, A.C., McClelland, J.L., Welbourne, S.R., Rogers, T.T., Ralph, M.A.: Why bilateral damage is worse than unilateral damage to the brain. Journal of Cognitive Neuroscience  \textbf{25},  2107--2123 (12 2013)

\bibitem{Selvaraju2016}
Selvaraju, R.R., Cogswell, M., Das, A., Vedantam, R., Parikh, D., Batra, D.: Grad-cam: Visual explanations from deep networks via gradient-based localization. In: 2017 IEEE International Conference on Computer Vision (ICCV) (2017)

\bibitem{Shillcock2001}
Shillcock, R., Monaghan, P.: The computational exploration of visual word recognition in a split model. Neural Computation  \textbf{13},  1171--1198 (5 2001)

\bibitem{simonyan_very_2015}
Simonyan, K., Zisserman, A.: Very deep convolutional networks for large-scale image recognition. arXiv preprint arXiv:1409.1556  (2015)

\bibitem{Tian2012}
Tian, J., Li, M., Chen, F., Kou, J.: Coevolutionary learning of neural network ensemble for complex classification tasks. Pattern Recognition  \textbf{45},  1373--1385 (4 2012)

\bibitem{Wang2013}
Wang, P., Cottrell, G.: A computational model of the development of hemispheric asymmetry of face processing. Proceedings of the Annual Meeting of the Cognitive Science Society  \textbf{35}, ~35 (2013)

\bibitem{weems_hemispheric_2004}
Weems, S.A., Reggia, J.A.: Hemispheric specialization and independence for word recognition: {A} comparison of three computational models. Brain and Language  \textbf{89}(3),  554--568 (Jun 2004), publisher: Academic Press

\end{thebibliography}

\end{document}